\documentclass[epj]{svjour}
\usepackage{graphics,graphicx}
\usepackage{amsmath,amsfonts}
\usepackage[square,numbers]{natbib}
\usepackage{appendix}
\usepackage{bm}
\usepackage{color}

\pdfoutput=1

\usepackage[normalem]{ulem}

\newcommand{\dpdS}[2]{\frac{\partial #1}{\partial #2}}

\newcommand{\dif}{\,\mathrm{d}}
\newcommand{\dpd}[2]{\frac{\partial {#1}}{\partial {#2}}}
\newcommand{\gradx}{\bm{\nabla}_{\! {\bm x}}}
\newcommand{\gradp}{\bm{\nabla}_{\! {\bm p}}}
\newcommand{\hatp}{\hat{\bm p}}

\newcommand{\hatz}{\hat{\bm z}}
\newcommand{\hatk}{\hat{\bm k}}

\begin{document}
\sloppy
\title{Influences of transversely-isotropic rheology and translational diffusion on the stability of active suspensions}
\author{Craig R. Holloway\inst{1} \and Gemma Cupples\inst{1} \and David J. Smith\inst{1,2} \and J. Edward F. Green\inst{3} \and Richard J. Clarke\inst{4} \and Rosemary J. Dyson\inst{1}
\thanks{\emph{e-mail:} R.J.Dyson@bham.ac.uk}%
}                     
%
%
\institute{School of Mathematics, University of Birmingham, B15 2TT, U.K. \and Institute for Metabolism and Systems Research, University of Birmingham, B15 2TT, U.K. \and School of Mathematical Sciences, University of Adelaide, Australia \and Department of Engineering Science, University of Auckland, New Zealand}
\date{Received: date / Revised version: date}
%
\abstract{
Suspensions of self-motile, elongated particles are a topic of significant current interest, exemplifying a form of `active matter'.
Examples include self-propelling bacteria, algae and sperm, and artificial swimmers.
Ericksen's model of a transversely-isotropic fluid [J. L. Ericksen, Colloid Polym. Sci. 173(2):117-122 (1960)] treats suspensions of non-motile particles as a continuum with an evolving preferred direction;
this model describes fibrous materials as diverse as extracellular matrix, textile tufts and plant cell walls.
Director-dependent effects are incorporated through a modified stress tensor with four viscosity-like parameters.
By making fundamental connections with recent models for active suspensions, we propose a modification to Ericksen's model, mainly the inclusion of self motility; this can be considered the simplest description of an oriented suspension including transversely-isotropic effects. 
Motivated by the fact that transversely-isotropic fluids exhibit modified flow stability, we conduct a linear stability analysis of two distinct cases, aligned and isotropic suspensions of elongated active particles. Novel aspects include the anisotropic rheology and translational diffusion.
In general anisotropic effects increase the instability of small perturbations, whilst translational diffusion stabilises a range of wave-directions and, in some cases, a finite range of wave-numbers, thus emphasising that both anisotropy and translational diffusion can have important effects in these systems.
} 

\authorrunning{Holloway \emph{et al.}}
\titlerunning{Influences of TI rheology and translational diffusion on the stability of active suspensions}
\maketitle


\section{Introduction}
Fluids containing suspensions of particles are found in numerous industrial and biological applications. Examples involving passive particles include (but are not limited to) solutions of DNA \cite{marrington2005validation}, fibrous proteins of the cytoskeleton \cite{dafforn2004protein,kruse2005generic}, synthetic bio-nanofibres \cite{mclachlan2013calculations}, extracellular matrix \cite{green2008extensional} and plant cell walls \cite{dyson2010fibre}.
Suspensions comprising self-propelling bacteria or other micro-organisms are termed \emph{active} \citep{saintillan2013active}; these suspensions exhibit phenomena such as collective behaviour \cite{pedley1990new,pedley1992hydrodynamic,hill2005bioconvection,saintillan2007orientational,lauga2009hydrodynamics,koch2011collective,hwang2014bioconvection} and, as observed recently, superfluidity \cite{lopez2015}. Collections of artificial swimmers may also exhibit the properties of active matter \cite{golestanian2005propulsion,dreyfus2005microscopic,yu2006experimental,howse2007self,patteson2016active}. 
In order to understand these phenomena, it is vital to develop tractable and accurate continuum theories that capture the essential physics of suspensions of self-motile particles.

In this paper we link active suspension models of solutions containing swimming microorganisms, such as those proposed by Pedley \& Kessler \cite{pedley1990new} and Simha \& Ramaswamy \cite{simha2002hydrodynamic}, with the mathematically simpler (inactive) transversely-isotropic fluid first described by Ericksen \cite{ericksen1960transversely}, commonly used to describe fibre-reinforced media \cite{green2008extensional,dyson2010fibre,dyson2015investigation,lee2005continuum,holloway2015linear}. 
Ericksen's model consists of mass and momentum conservation equations together with an evolution equation for the fibre director field. The stress tensor depends on the fibre orientation and linearly on the rate of strain; it takes the simplest form that satisfies the required invariances.

By linking these models of active and inactive suspensions, we propose a modification to the fibre evolution equation, of the transversely-isotropic model, which allows for the inclusion of swimming particles. Alongside this, the connection between the two models reveals the importance of non-isotropic terms in the stress tensor, when the suspended particles are elongated; these terms are known to influence the dynamics of fibre-laden flows \citep{holloway2015linear,rogers1989squeezing,hull1992,spencer1997}.

Motivated by these anisotropic terms, we then analyse the linear stability of suspensions of elongated particles, with zero imposed background flow, in two distinct cases, following the analysis undertaken in ref.\ \citep{saintillan2008instabilities}; when the particles are nearly aligned and when the suspension is isotropic, \emph{i.e.}\ when particles have nearly zero dispersion at each point in space, and when the particles are perfectly randomly orientated respectively.
We extend the analysis of ref.\ \citep{saintillan2008instabilities} by including translational diffusion in the analysis of the aligned case, and anisotropic effects in both cases.
 We find that the inclusion of translational diffusion does have a stabilising effect in the aligned case, however, the magnitude of this effect is not uniform for all wave-directions. In general, the stabilising effect is strongest for wave-directions near perpendicular to the aligned direction and weakest for near parallel. The importance of the non-isotropic stress has been identified in the transversely-isotropic research literature \citep{green2008extensional,holloway2015linear,cupples2017viscous}.
 
 The structure of this paper is as follows: in section \ref{GoverningEquationsForAnActiveSuspension} we propose the governing equations for an active suspension of elongated particles; in section \ref{transversely} we show that the active description of a uniformly-distributed, perfectly-aligned suspension is equivalent to Ericksen's model \citep{ericksen1960transversely} with a modified director evolution equation; in section \ref{stability} we analyse the linear stability of the two distinct cases of aligned and isotropic suspensions of elongated particles, taking into account transversely-isotropic rheology and translational diffusion, before giving a brief summary of our findings in section \ref{summary}.

\section{Governing equations for an Active Suspension}\label{GoverningEquationsForAnActiveSuspension}
\begin{figure}
\centering
\includegraphics[width=.3\textwidth]{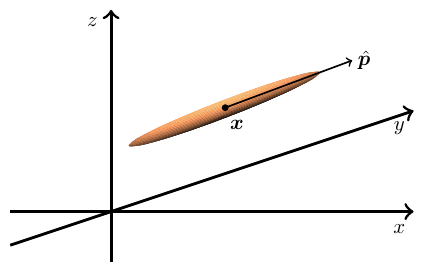}
\caption{A schematic diagram showing the coordinate system used to model the particle distribution function $N^*$. The particle's position in space is given by the vector ${\bm x}^*=(x^*,y^*,z^*)$ and its orientation is given by the unit vector $\hatp$.}\label{fig:1}
\end{figure}

Consider a collection of particles suspended in a viscous, Newtonian fluid. The density of particles is sufficiently dilute that particles do not interact directly, only through their influence on the fluid. 
Each particle is modelled as a prolate spheroid with major axis $r_1^*$, minor axis $r_2^*$, aspect ratio $\Gamma=r_1^*/r_2^*$, and shape parameter $\alpha_0=(\Gamma^2-1)/(1+\Gamma^2)$.
The particle number density in physical and orientation space is denoted $N^*({\bm x}^*,\hatp,t^*)$ where ${\bm x}^*$ denotes the particle position, $\hatp$ is orientation and $t^*$ is time \citep{doi1988theory}, visualised in Fig. \ref{fig:1}.
This function is normalised such that
\begin{equation}
\frac{1}{V^*} \int_{V^*} \int_S N^*({\bm x}^*,\hatp,t^*) \dif \hatp \dif {\bm x}^* = n_d^*, \label{eq:norm*}
\end{equation}
where $V^*$ is the volume of the spatial domain, $S$ is the surface of the unit sphere in orientation space and $n_d^*$ is the mean number density of particles in the suspension. 
The local particle director and concentration fields ${{\bm a} = \langle \hatp \rangle}$ and $c^*$ are defined such that
\begin{eqnarray}
{\bm a} = \langle \hatp \rangle ({\bm x}^*,t^*) &=& \frac{1}{c^*({\bm x}^*,t^*)} \int_S \hatp \, N^*({\bm x}^*,\hatp,t^*) \dif \hatp, \label{eq:av*}\\
c^*({\bm x}^*,t^*) &=& \int_S N^*({\bm x}^*,\hatp,t^*) \dif \hatp. \label{eq:n*}
\end{eqnarray}
The bracket operator is defined over other quantities similarly.

The particle distribution function is assumed to be governed by a Fokker-Planck equation \cite{saintillan2013active}, giving a conservation law for $N^*$:
\begin{equation}
\dpdS{N^*}{t^*} + \gradx^* \cdot \left( {\bm U}^* N^* \right) + \gradp \cdot \left( \bm\Omega^* N^* \right) = 0,\label{eq:N*}
\end{equation}
where $\gradx^*$ denotes the gradient operator in physical space and $\bm\gradp$ denotes the gradient operator on the unit sphere in orientation space.
The particle translational velocity ${\bm U}^*$ is represented by the linear combination of the particle swimming velocity relative to a background flow $U_s^* \hatp$, the local fluid velocity ${\bm u}^*$ and translational diffusion (with diffusivity constant $D_T^*$) \citep{saintillan2007orientational,saintillan2008instabilities}:
\begin{equation}
{\bm U}^* = U_s^* \hatp + {\bm u}^* -D_T^* \gradx^* \left( \ln N^* \right). \label{eq:xdot*}
\end{equation}
In this paper we will treat $D_T^*$ as a free parameter, as opposed to relating it to random properties of swimming via either a phenomenological argument or Taylor dispersion theory \citep{hill2002}.

Jeffery's equation \citep{jeffery1922motion} models the angular velocity of the particle, in the absence of rotational diffusion, as
\begin{equation}
\bm\Omega^* = \left( {\bm I} - \hatp \, \hatp \right) \cdot \left[ \left( \alpha_0 {\bm e}^* + \bm\omega^* \right) \cdot \hatp \right], \label{eq:f*}
\end{equation}
where we denote the rate-of-strain tensor ${\bm e}^* = ( \gradx^* {\bm u}^* + \gradx^* {\bm u}^{*T} )/2$, the vorticity tensor $\bm\omega^* = ( \gradx^* {\bm u}^* - \gradx^* {\bm u}^{*T} )/2$ and the identity tensor ${\bm I}$.

Finally, the fluid velocity ${\bm u}^*({\bm x}^*,t^*)$ is governed by the Cauchy momentum equations
\begin{eqnarray}
\rho^* \left( \dpd{{\bm u}^*}{t^*} + \left( \bm{u}^* \cdot \gradx^* \right) \bm{u}^* \right)&=& \gradx^* \cdot \bm\sigma^* ,  \nonumber\\
\gradx^* \cdot {\bm u}^* &=& 0,  \label{eq:flow*}
\end{eqnarray}
where $\rho^*$ is the fluid density and $\bm\sigma^*$ is the stress tensor, which must be prescribed by a constitutive law.
Although the fluid containing the particles is assumed to be Newtonian and isotropic, the presence of the particles will induce anisotropic behaviour.

Most models currently found in the literature take account of the isotropic ($\bm\sigma_I^*$) and active ($\bm\sigma_S^*$) contributions to the stress, but neglect the interaction of the particle with the surrounding fluid ($\bm\sigma_P^*$).
We therefore follow Pedley \& Kessler \cite{pedley1990new} and take an expression for the stress tensor of the form
\begin{equation}
\bm\sigma^* = \bm\sigma_I^* + \bm\sigma_S^* + \bm\sigma_P^*.  \label{eq:stress*}
\end{equation}

The isotropic component takes the form
\begin{equation}
\bm\sigma_I^* = -\bar{p}^* \, {\bm I} + 2 \, \bar{\mu}^* \, {\bm e}^*,
\end{equation}
where $\bar{p}^*$ is the hydrostatic pressure and $\bar{\mu}^*$ is the solvent viscosity.

Active behaviour of force-free Stokesian swimmers is modelled by an equal and opposite propulsive force/drag pair acting along, and infinitesimally displaced in, the $\pm \hatp$ direction.
By differentiating the Oseen tensor the resulting flow field is of the form of a symmetric stokes dipole (`stresslet') with tensorial strength proportional to ($\hatp \, \hatp - {\bm I}/3$).
As shown by Batchelor \cite{batchelor1970stress}, this flow contributes proportionally to the bulk stress.
Averaging over orientation space, the contribution to the stress due to active swimming is therefore of the form 
\begin{equation}
\bm\sigma_S^* = \alpha_1^* \, c^*  \, \left\langle  \hatp \, \hatp - \frac{\bm I}{3} \right\rangle, \label{eq:sigmas*}
\end{equation}
where $\alpha_1^*$ is a parameter, which could be positive (puller) or negative (pusher), quantifying the active stresslet strength. 

The components of the stress tensor that arise from the presence of suspended particles in the solvent take the form \citep{batchelor1970stress,hinch1972effect}
\begin{eqnarray}
\bm\sigma_P^* &=& 4 \, \bar{\mu}^* \, V_c^* \Biggl[ \alpha_2 {\bm e}^* : \displaystyle\int_S {\hatp \, \hatp  \, \hatp  \, \hatp} \, N^* \dif \hatp \nonumber \\
& +& \alpha_3 \left( {\bm e}^* \cdot \displaystyle\int_S {\hatp \,  \hatp} \, N^* \dif \hatp + \displaystyle\int_S {\hatp \, \hatp} \, N^* \dif \hatp \cdot {\bm e}^* \right) \nonumber \\
&  +& \alpha_4 {\bm e}^* \displaystyle\int_S N^* \dif \hatp + \alpha_5 \, {\bm I} \,  {\bm e}^* : \displaystyle\int_S {\hatp \,\hatp} \, N^* \dif \hatp \Biggr], \label{eq:stress_last}
\end{eqnarray}
where $V_c^*=4 \pi  r_1^{*} r_2^{*2}/3$ is the particle volume and $\alpha_i$ ($i=2\dots5$) are constants. The terms in $\alpha_2$ and $\alpha_5$ can be identified with equation $(17)$ in the extensional flow study of ref.\ \citep{saintillan2010extensional}. Note that while $\alpha_1^*$ is dimensional, $\alpha_2$\,--\,$\alpha_5$ are dimensionless.

The full model thus consists of a normalisation condition for $N^*$ \eqref{eq:norm*}, where $N^*$ is governed by the Fokker-Planck equation \eqref{eq:N*}, with fluxes \eqref{eq:xdot*} and \eqref{eq:f*}.
The fluid velocity obeys conservation of mass and momentum \eqref{eq:flow*}, with a constitutive relation for stress given by equations \eqref{eq:stress*}--\eqref{eq:stress_last}.

\section{Transversely-Isotropic fluid}\label{transversely}
In this section we show how the model, proposed in section \ref{GoverningEquationsForAnActiveSuspension}, may be related to the model of a transversely-isotropic fluid, proposed by Ericksen \citep{ericksen1960transversely}.
Consider a uniform suspension $c^*({\bm x}^*,t^*)=n_d^*$ which is perfectly aligned, with director field ${\bm a}({\bm x}^*,t^*)$, and where angular and translational diffusion are neglected; the particle distribution function is then of the form
\begin{equation}
N^*\left({\bm x}^*,\hatp,t^*\right) = n_d^* \, \delta \left( \hatp - {\bm a}  \right),
\end{equation}
where $\delta$ denotes the Dirac delta function \citep{saintillan2008instabilities}.
In this section only we set the translational diffusion coefficient to zero; in section \ref{stability} we will reintroduce translational diffusion to consider its effect on stability.
In this case we need only consider how the average direction of the particles ${\bm a}$ evolves, and not the full distribution function $N^*$.
The evolution equation for ${\bm a}$ is derived by multiplying equation \eqref{eq:N*} by $\hatp$ and integrating over $\hatp$, to give \citep{saintillan2008instabilities}
\begin{eqnarray}
\begin{split}
\dpd{\bm a}{t^*} + \left( U_s^* \, {\bm a} + {\bm u}^* \right) \cdot & \gradx^* \, {\bm a} - \bm\omega^* \cdot {\bm a} \\
&= \alpha_0 \left( {\bm e}^* \cdot  {\bm a} - {\bm e}^* : {\bm a} \, {\bm a} \, {\bm a} \right).  \label{eq:av_a}
\end{split}
\end{eqnarray}
The fibre evolution equation of Ericksen \citep{ericksen1960transversely} for a passive transversely-isotropic fluid can then be recovered by setting the swimming speed to zero ($U_s^*=0$).

The governing equations for the background flow (equations \eqref{eq:flow*}) remain unchanged, however the stress tensor is now given by
\begin{eqnarray}
\bm\sigma^*=& - p^* \, {\bm I} + 2 \mu^* \, {\bm e}^* + \mu_1^* \, {\bm a}\, {\bm a}\, + \mu_2^* \, {\bm a}\, {\bm a}\, {\bm a}\, {\bm a} : {\bm e}^* \nonumber \\
& + 2 \mu_3^* \left( {\bm a}\, {\bm a}\, \cdot {\bm e}^* + {\bm e}^* \cdot {\bm a}\, {\bm a}\,\right), \label{eq:stress}
\end{eqnarray}
where the pressure has been modified such that,
\begin{eqnarray}
p^* & =&\bar{p}^* + \frac{n_d^* \alpha_1^*}{3} -4 \, \bar{\mu}^* \, \phi \, \alpha_5 \,  {\bm a}  {\bm a}  : {\bm e}^*,
\end{eqnarray}
and the viscosity-like parameters are given by
\begin{align}
\mu^* &= \bar{\mu}^* \left( 1 + 2 \, \phi \, \alpha_4  \right), \nonumber &
\mu_1^* &=n_d^* \, \alpha_1^*, \nonumber \\
\mu_2^* &= 4 \, \bar{\mu}^* \,  \phi \,  \alpha_2, \nonumber&
\mu_3^* &= 2 \, \bar{\mu}^* \, \phi \, \alpha_3.
\end{align}
The non-dimensional parameter $\phi= n_d^*  V_c^*$ denotes the volume fraction of the particles.
These parameters may be interpreted in turn as follows: $\mu^*$ is the shear viscosity in the direction transverse to the particles, this is equivalent to the solvent viscosity enhanced by the volume fraction of particles \cite{dyson2010fibre};
$\mu_1^*$ implies the existence of a stress in the fluid even if it is instantaneously at rest generated via an active stresslet; $\mu_2^*$ and $\mu_3^*$ are the anisotropic extensional and shear viscosities respectively due to the presence of the particles \cite{green2008extensional,dyson2010fibre,rogers1989squeezing,holloway2015linear,dyson2015investigation}.

When the suspension is dilute the non-dimensional parameters $\mu=1 + 2 \, \phi \,\alpha_4$, $\alpha_2$ and $\alpha_3$ may be approximated from Jeffery \citep{jeffery1922motion} (via \citep{pedley1990new,batchelor1970stress,leal1971effect,brenner1972rheology,kim2013microhydrodynamics}); these approximations are shown in appendix \ref{app:Jeffery}.

We have therefore recovered, from a description of an active suspension of aligned elongated particles, the model for a transversely-isotropic fluid proposed by Ericksen \cite{ericksen1960transversely}, with a modification to the fibre evolution equation to account for the swimming velocity. When the fibre evolution equation is not modified, Ericksen's transversely-isotropic model corresponds to a non-swimming `shaker' suspension \citep{saintillan2015theory}, and as such offers a simple model of, for example, the dynamics of suspensions of microtubule bundles that extend in length due to motor-protein activity \citep{sanchez2012spontaneous}. By analogy with recent findings in the transversely-isotropic field \citep{dyson2010fibre,dyson2015investigation,holloway2015linear,cupples2017viscous}, this correspondence suggests that the transversely-isotropic stress will have a significant impact on the dynamics of suspensions of active elongated particles, via the rheological parameters $\alpha_2$ and $\alpha_3$.

In the next section we examine the role of these transversely-isotropic effects, along with translational diffusion, in the stability analysis of both aligned and isotropic suspensions of elongated particles.

\section{Stability of nearly-aligned and isotropic suspensions}\label{stability}
\begin{figure}[h]
\centering
\includegraphics[width=.5\textwidth]{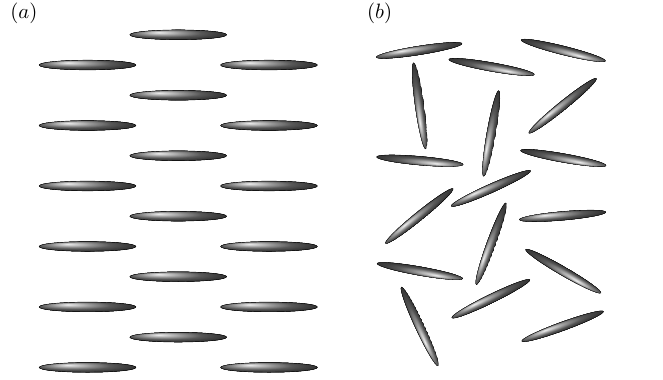}
\caption{Schematic diagrams of (a) an aligned suspension and (b) an isotropic suspension of rod-like particles}\label{AIfig:aligned}
\end{figure}

In this section we examine the linear stability of suspensions of elongated particles, with zero background flow, in two distinct cases, when $(1)$ the particles are nearly aligned and $(2)$ the suspension is isotropic (a schematic diagram is given in Fig. \ref{AIfig:aligned})

We adopt the model described in section \ref{GoverningEquationsForAnActiveSuspension}. 
To derive a governing equation for the concentration field $c^*({\bm x}^*,t^*)$, equation \eqref{eq:N*} is integrated with respect to $\hatp$  over orientation space; after substituting for the translational flux velocity $\bm{U}^*$ (equation \eqref{eq:xdot*}) this yields \citep{saintillan2008instabilities}
\begin{align}
\dpdS{c^*}{t^*} + {\bm u}^* \cdot \gradx^* c^* &= D_T^* \gradx^{*2} c^* - U_s^* \gradx^* \cdot \left( c^* {\bm a} \right).\label{AIeq:conc*}
\end{align}
 Equation \eqref{AIeq:conc*} is an advection-diffusion equation for the local concentration field $c^*$, with a source term {$-U_s^* \gradx^* \cdot \nobreak ( c^* {\bm a} )$}.

\subsection{Non-dimensionalisation}
The governing equations are made dimensionless using the following scaling \citep{saintillan2008instabilities}:
\begin{align}
{\bm u}^* &= U_s^* {\bm u}, & {\bm x}^* &= \frac{\bm x}{n_d^* r_1^{*2}}, & t^* &= \frac{t}{n_d^* r_1^{*2} U_s^*}, \nonumber \\ \bar{p}^* &= \bar{\mu}^* U_s^* n_d^* r_1^{*2} p, & N^* &= n_d^* N, & c^* &= n_d^* c,
\end{align}
where $n_d^*$ is the mean number density of particles in the suspension. Note that $n_d^* r_1^{*2} = V_e^*/V^* r_1^*$, where $V_e^*=M r_1^{*3}$ is the effective volume taken up by the total number of swimming particles ($M$). Choosing these scalings to non-dimensionalise the model is only consistent when the background flow is approximately zero.

The particle distribution function (equation \eqref{eq:norm*}) is now normalised as 
\begin{align}
\frac{1}{V} \int_{V} \int_S N({\bm x},\hatp,t) \dif \hatp \dif {\bm x} = 1, \label{eq:norm}
\end{align}
and the governing equation (equation \eqref{eq:N*}) becomes
\begin{align}
\dpdS{N}{t} &= - \gradx \cdot \left( {\bm U} N \right) - \gradp \cdot \left( \bm\Omega N \right),\label{eq:N} 
\end{align}
where the non-dimensional translational and rotational velocities of the particles (equation \eqref{eq:xdot*} and \eqref{eq:f*}) are given by
\begin{align}
{\bm U} &= \hatp + {\bm u} - \frac{\phi}{\overline{P}\!_e}  \gradx \left( \ln N \right),\label{eq:U}  \\
\bm\Omega &= \left( {\bm I} - \hatp \, \hatp \right) \cdot \left[ \left( \alpha_0 {\bm e} + \bm\omega \right) \cdot \hatp  \right]. \label{eq:Omega}
\end{align}
Here ${\bm e}= (\gradx {\bm u} + \gradx {\bm u}^T)/2$ is the rate-of-strain tensor, $\bm\omega = (\gradx {\bm u} - \gradx {\bm u}^T)/2$ is the vorticity tensor, $\phi = n_d^* V_c^*$ is the volume fraction of particles in solution, and ${\overline{P}\!_e = 3 U_s^* r_1^* / 4 \pi D_T^* \Gamma^2}$ is the modified P\'{e}clet number, which is the dimensionless ratio of the convection time scale to the diffusion time scale \citep{mclachlan2013calculations}, divided by the square of the aspect ratio. 

The local particle director and concentration fields (equations \eqref{eq:av*} and \eqref{eq:n*}) are 
\begin{align}
{\bm a}({\bm x},t) = \langle \hatp \rangle &= \frac{1}{c({\bm x},t)} \int_S \hatp N({\bm x}, \hatp,t) \dif \hatp, \\
c({\bm x},t) &= \int_S N({\bm x},\hatp,t) \dif \hatp,
\end{align}
where the concentration field (equation \eqref{AIeq:conc*}) is governed by 
\begin{align}
\dpd{c}{t} + \left( {\bm u} \cdot \gradx  \right)c &= \frac{\phi}{\overline{P}\!_e} \nabla_x^2 c  - \gradx \cdot \left( c {\bm a} \right). \label{conc}
\end{align}
Finally, in the zero Reynolds number limit the momentum and continuity equations \eqref{eq:flow*} simplify as
\begin{align}
\gradx \cdot  \bm\sigma &= \bm0, & \gradx \cdot  {\bm u} &= 0, \label{eq:flow}
\end{align}
where the constitutive relation for stress \eqref{eq:stress_last} becomes
\begin{align}
\begin{split}
\bm\sigma =& -p \, {\bm I} + 2 \, {\bm e} +  \alpha_1 \int_S \left( \hatp \, \hatp - \frac{\bm I}{3} \right) N \dif \hatp \\
&+ 4 \phi c  \Bigg\{ \alpha_2 {\bm e} : \int_S \hatp \, \hatp \, \hatp \, \hatp \, N \dif \hatp \\ &+ \alpha_3 \left( {\bm e} \cdot \int_S \hatp \, \hatp \, N \dif \hatp + \int_S \hatp \, \hatp \, N \dif \hatp \cdot {\bm e} \right) \\
&+ \alpha_4 {\bm e} \int_S N \dif \hatp + \alpha_5 {\bm e} : \int_S \hatp \, \hatp \, N \dif \hatp \, {\bm I} \Bigg\}.
\end{split} \label{stress}
\end{align}
Here $\alpha_1 = \alpha_1^*/\bar{\mu}^* U_s^* r_1^{*2}$ is the non-dimensional stresslet strength.

\subsection{Stability of a nearly-aligned suspension}
First consider the case when the particles are perfectly aligned at each point ${\bm x}$ and the imposed background flow is zero to leading order. 
This situation may arise after particles have been orientated by a background flow, followed by the flow being instantaneously turned off.

For a nearly-aligned suspension the distribution function takes the form 
\begin{align}
N({\bm x}, \hatp,t) &= c ({\bm x},t) \delta \left( \hatp - {\bm a} ({\bm x},t) \right), \label{eq:529}
\end{align}
where $\delta$ denotes the Dirac delta function \cite{saintillan2008instabilities}. This is similar to the form of the distribution function chosen in section \ref{transversely}, but with non-uniform concentration.
In this case we may reduce the evolution equation for the full distribution function \eqref{eq:N} to a pair of equations for the concentration and director fields.
These latter equations now only have $\bm{x}$ and $t$ as independent variables, rather than $\bm{x}$, $\hatp$ and $t$, reducing the dimensionality of the problem.
The equation for concentration (equation \eqref{conc}) is given by
\begin{align}
\dpd{c}{t} + \gradx \cdot \left[ \left( {\bm a} + {\bm u} \right) c \right] &= \frac{\phi}{\overline{P}\!_e} \nabla_x^2 c. \label{eq:align_conc} 
\end{align}
The evolution equation for the director field $\bm{a}$ is derived by, multiplying equation \eqref{eq:N} by $\hatp$ and integrating with respect to $\hatp$, yielding \cite{saintillan2008instabilities},
\begin{align}
\dpd{\bm{a}}{t} &= - \left( \bm{a} + \bm{u} \right) \cdot \gradx \bm{a} +  \left( \bm{I} - \bm{a} \, \bm{a} \right) \cdot \left[ \left( \alpha_0 \bm{e} + \bm{\omega} \right) \cdot \bm{a} \right], \label{eq:align_N}
\end{align}
where the concentration is assumed non-zero everywhere.

The governing equations for the fluid velocity remain unchanged (equations \eqref{eq:flow}), however the stress tensor is now given by
\begin{align}
\begin{split}
\bm\sigma =& -p \, {\bm I} + 2 \, {\bm e} +  \alpha_1 \, c(\bm{x},t) \left( {\bm a} \, {\bm a} - \frac{\bm I}{3} \right)  \\& + 4 \, \phi \, c({\bm x},t)  \Bigg( \alpha_2  {\bm e} : {\bm a} \, {\bm a} \, {\bm a} \, {\bm a}   + \alpha_3 \left( {\bm e} \cdot {\bm a} \, {\bm a} + {\bm a} \, {\bm a} \cdot {\bm e} \right) \\ 
& \hspace{4.7cm}+ \alpha_4 {\bm e} + \alpha_5 {\bm e} : {\bm a} \, {\bm a} \, {\bm I} \Bigg). \label{eq:align_stress}
\end{split}
\end{align}
In line with the transversely-isotropic fluid literature we define $\alpha_2$ and $\alpha_3$ as the anisotropic extensional and shear viscosities respectively, and note that $4 \, \phi \, c \, \alpha_4$ and $4 \, \phi \, c \, \alpha_5 \, \bm{e} : \bm{a} \, \bm{a}$ may be combined with the solvent viscosity and hydrostatic pressure respectively.

The model consists of equations for the evolution of concentration  \eqref{eq:align_conc} and director \eqref{eq:align_N} fields of the particles, as well as conservation of mass and momentum statements for the fluid velocity \eqref{eq:flow}, with the constitutive relation for stress \eqref{eq:align_stress}. 

As no external body forces act upon the fluid flow, we may analyse the linear stability of a suspension of particles aligned in the $\hat{\bm z}$-direction (\emph{i.e.}\ ${\bm a}^{(0)}=\hatz$) without loss of generality \citep{saintillan2008instabilities}.
A base state exists when the fluid is motionless (${\bm u}^{(0)}=\bm0$), the particles are uniformly distributed ($c^{(0)}=1$), and the pressure is constant ($p^{(0)} = p_0$, where $p_0$ is some arbitrary pressure).
We consider the stability of this state via the perturbation
\begin{align}
c({\bm x},t) &= 1 + \varepsilon c^{(1)}({\bm x},t) + \mathcal{O} \left( \varepsilon^2 \right) ,\nonumber \\ {\bm a}(\bm{x},t) &= \hat{\bm z} + \varepsilon {\bm a}^{(1)} ({\bm x},t), + \mathcal{O} \left( \varepsilon^2 \right), \nonumber \\
{\bm u}({\bm x},t) &= \varepsilon {\bm u}^{(1)}({\bm x},t) + \mathcal{O} \left( \varepsilon^2 \right), \nonumber \\ 
p({\bm x},t) &= p_0 + \varepsilon p^{(1)}({\bm x},t) + \mathcal{O} \left( \varepsilon^2 \right), \label{AIeq:aper}
\end{align}
where $|\varepsilon|\ll1$. We require ${\bm a}^{(1)} \cdot \hat{\bm z} =0$, so that ${\bm a}$ remains a unit vector to order $\varepsilon^2$.

Expanding equations \eqref{eq:align_conc} and \eqref{eq:align_N}, and retaining only terms of order $\varepsilon$ only, we find \citep{saintillan2008instabilities}
\begin{align}
\dpd{c^{(1)}}{t} + \hat{\bm z} \cdot \gradx c^{(1)}  &= \frac{\phi}{\overline{P}\!_e} \nabla_x^2 c^{(1)} - \gradx \cdot {\bm a}^{(1)}, \label{LazyConc1}\\
\dpd{{\bm a}^{(1)}}{t} + \hat{\bm z} \cdot \gradx {\bm a}^{(1)} &= \left( {\bm I} - \hat{\bm z} \, \hat{\bm z} \right) \cdot \left( \alpha_0 {\bm e}^{(1)} + \bm\omega^{(1)} \right) \cdot \hat{\bm z}, \label{LazyAlign1}
\end{align}
where ${\bm e}^{(1)} = (\gradx {\bm u}^{(1)} + \gradx {\bm u}^{(1)T})/2$ and $\bm\omega^{(1)} = (\gradx {\bm u}^{(1)} - \gradx {\bm u}^{(1)T})/2$.
The momentum equations are given at order $\epsilon$ by
\begin{align}
-\mu \nabla_{\bm x}^2 {\bm u}^{(1)} + \gradx q^{(1)} &= \gradx \cdot \bm\sigma^{(1)}, &
\gradx \cdot {\bm u}^{(1)} &= 0, \label{AIeq:linear_align_flow}
\end{align}
where $\mu = 1 + 2 \phi \alpha_4$ is the enhanced shear viscosity due to the presence of the particles, the effective pressure is $q^{(1)} = p^{(1)} - \alpha_1 c^{(1)} /3 - 4 \phi \alpha_5 {\bm e}^{(1)} : \hat{\bm z} \hat{\bm z}$, and the constitutive relation for stress is given by
\begin{align}
\begin{split}
\bm\sigma^{(1)} =& \, \alpha_1 \left( {\bm a}^{(1)} \hat{\bm z} + \hat{\bm z} {\bm a}^{(1)} + c^{(1)} \hat{\bm z} \hat{\bm z} \right) \\
& + 4 \phi \left\{ \alpha_2 {\bm e}^{(1)} : \hat{\bm z} \hat{\bm z} \hat{\bm z} \hat{\bm z} + \alpha_3 \left( {\bm e}^{(1)} \cdot \hat{\bm z} \hat{\bm z} + \hat{\bm z} \hat{\bm z} \cdot {\bm e}^{(1)} \right) \right \}. \label{stress1}
\end{split}
\end{align}
When the suspension is dilute the parameters $\mu,$ $\alpha_2$ and $\alpha_3$ may be approximated from Jeffery \citep{jeffery1922motion} (via \citep{pedley1990new,batchelor1970stress,leal1971effect,brenner1972rheology,kim2013microhydrodynamics}), see appendix \ref{app:Jeffery}.

We seek plane-wave solutions of the form 
\begin{align}
c^{(1)} &= c' (\bm{k}) e^{i{\bm k}\cdot {\bm x} + s t},&  {\bm a}^{(1)}&= {\bm a}' (\bm{k}) e^{i{\bm k}\cdot {\bm x} + s t},  \nonumber \\ q^{(1)} &= q' (\bm{k}) e^{i{\bm k}\cdot {\bm x} + s t}, & {\bm u}^{(1)}&=  {\bm u}'(\bm{k}) e^{i{\bm k}\cdot {\bm x} + s t}, \\
{\bm e}^{(1)}&=  {\bm e}'(\bm{k}) e^{i{\bm k}\cdot {\bm x} + s t}, & {\bm \sigma}^{(1)}&=  {\bm \sigma}'(\bm{k}) e^{i{\bm k}\cdot {\bm x} + s t},
\end{align}
where ${\bm k}$ is the wave-vector and $s$ the growth rate. 
Under this ansatz the equations for the first order concentration and alignment field (equations \eqref{LazyConc1} and \eqref{LazyAlign1}) are given by
\begin{align}
\left( s + i \hatz \cdot \bm{k} + \frac{\phi}{\overline{P}\!_e} \nabla_x^2 \right) c' &= - i \bm{k} \cdot \bm{a}', \label{aligned:conc} \\
\left( s + i \hatz \cdot \bm{k} \right) \bm{a}' &= \left( \bm{I} - \hatz \, \hatz \right) \cdot \left( \alpha_0 \bm{e}' + \bm\omega' \right) \cdot \hatz, \label{aligned:alignment}
\end{align}
where $\bm{e}' = i (\bm{k \, u}' + \bm{u}' \, \bm{k} )/2$ and $\bm\omega' = i (\bm{k \, u}' - \bm{u}' \, \bm{k})/2$, whilst the conservation of mass and momentum equations become \citep{saintillan2008instabilities}
\begin{align}
\mu k^2 \bm{u}' + i \bm{k} q' &= i \bm{k} \cdot \bm{\sigma}', & \bm{k} \cdot \bm{u}' &= 0. \label{aligned:FlowWithQ}
\end{align}
The constitutive relation for stress is now given by 
\begin{eqnarray}
	\bm\sigma' &=& \, \alpha_1 \left( {\bm a}' \hat{\bm z} + \hat{\bm z} {\bm a}' + c' \hat{\bm z} \hat{\bm z} \right) \nonumber \\
&& + 4 \phi \left\{ \alpha_2 {\bm e}' : \hat{\bm z} \hat{\bm z} \hat{\bm z} \hat{\bm z} + \alpha_3 \left( {\bm e}' \cdot \hat{\bm z} \hat{\bm z} + \hat{\bm z} \hat{\bm z} \cdot {\bm e}' \right) \right \}. \label{aligned:stress}
\end{eqnarray}

Utilising the constitutive relation for stress \eqref{stress1} and eliminating the effective pressure $q'$, allows the conservation of momentum statement \eqref{aligned:FlowWithQ} to be written as
\begin{align}
	\begin{split}
\Bigg( 1 &+ \frac{ 2 \phi \alpha_3 }{\mu} \left( \hat{\bm k} \cdot \hat{\bm z} \right)^2 \Bigg) {\bm u}' \\
+& \frac{2 \phi}{\mu} \left( {\bm u}' \cdot \hat{\bm z} \right) \left( 2 \alpha_2 \left( \hat{\bm k} \cdot \hat{\bm z} \right)^2 + \alpha_3 \right) \left( \hat{\bm z} - \left( \hat{\bm k} \cdot \hat{\bm z} \right) \hat{\bm k} \right) \\
&= \frac{ i \alpha_1}{\mu k^2} \left( {\bm I} - \hat{\bm k} \hat{\bm k} \right) \cdot \left( {\bm a}' \hat{\bm z} + \hat{\bm z} \, {\bm a}' + c' \hat{\bm z} \, \hat{\bm z} \right) \cdot {\bm k}. \label{aligned:velocity}
\end{split}
\end{align}

Similarly to the case identified by ref.\ \citep{saintillan2007orientational,saintillan2008instabilities}, the velocity is only non-zero if the wave-vector ${\bm k}$ lies in the $(\hat{\bm z},{\bm a}')$ plane, therefore we may
assume without loss of generality that ${\bm k}$ lies in this plane and define $\theta$ as the angle between ${\bm k}$ and $\hat{\bm z}$, \emph{i.e.}\ ${\bm k} = k ( \cos \theta \, \hat{\bm z} + \sin \theta \, {\bm a}'/a')$ (where $a'=|{\bm a}'|$). Assuming this form for the wave-vector, the concentration and alignment equations become
\begin{align}
\lambda' c' &= - i \, k \, a' \, \sin \theta, \label{aligned:ConWave} \\
\lambda &= \frac{i}{2} \left( \left( \alpha_0 + 1 \right) u_a k \cos \theta + \left( \alpha_0 -1 \right) u_z k \sin \theta \right),\label{aligned:AliWave}
\end{align}
where $\lambda' = s + i k \cos \theta + \phi  k^2 / \overline{P}\!_e$ and $\lambda = s + i k \cos \theta$, chosen for notational convenience.
Here the components of velocity are given by
\begin{align}
\begin{split}
u_a' =& \frac{i \alpha_1 a'  \left( \mu + 2 \phi \alpha_3 \cos 2 \theta \right) }{k \lambda' (\mu + \phi ( \alpha_2 \sin^2 2 \theta + 2 \alpha_3 ))(\mu + 2 \phi \alpha_3 \cos^2 \theta)}  \\
&\times \left( \lambda' \cos \theta \cos 2 \theta + i k \sin^2 \theta \cos^2 \theta \right)  ,
\end{split} \label{aligned:ua} \\
 u_z' =& \frac{- i \alpha_1 a' \left(  \lambda' \sin \theta \cos 2 \theta + ik  \cos \theta \sin^3 \theta \right)}{k \lambda' \left(\mu+ \phi \left( \alpha_2 \sin^2 2 \theta + 2 \alpha_3 \right) \right)}, \label{aligned:uz}
\end{align}
where we have made use of equation \eqref{aligned:ConWave} to eliminate $c'$.

Substitution of the velocity components \eqref{aligned:ua} and \eqref{aligned:uz} into equation \eqref{aligned:AliWave} leads to the dispersion relation
\begin{align}
\lambda \, \lambda' - f \left( \theta \right) \left( \lambda' \cos 2 \theta +i k \sin^2 \theta \cos \theta \right) &= 0,\label{46}
\end{align}
where 
\begin{align}
f (\theta) &= - \frac{\alpha_1}{2} \left( A_1 \left( \alpha_0 + 1 \right) \cos^2 \theta - A_2 \left( \alpha_0 -1 \right) \sin^2 \theta \right),	 \label{eq:47}\\
A_1 &= \frac{ \mu + 2 \phi \alpha_3 \cos 2 \theta }{(\mu + \phi ( \alpha_2 \sin^2 2 \theta + 2 \alpha_3))(\mu+ 2 \phi \alpha_3 \cos^2 \theta)}, \label{eq:48} \\
A_2 &= \frac{1}{\mu + \phi (  \alpha_2 \sin^2 2 \theta + 2 \alpha_3)}. \label{eq:49}
\end{align}
Equation \eqref{46} is an eigenvalue problem for the growth rate $s$ (via $\lambda$ and $\lambda'$), the solution for which is obtained as 

\begin{align}
\begin{split}
	s_\pm =& -i k \cos \theta - \frac{\phi k^2}{\overline{P}\!_e} +  \frac{1}{2} \left( \frac{\phi k^2}{\overline{P}\!_e} + f(\theta) \cos 2 \theta \right) \\
	&\times \left[ 1 \pm \left( 1 + \frac{ 2 i k f(\theta) \sin \theta \sin 2 \theta}{\left( \phi k^2/\overline{P}\!_e + f(\theta) \cos 2 \theta \right)^2 } \right)^{1/2} \right].
	\end{split}\label{dispersion}
\end{align}
We note the solution of the eigenvalue problem derived by Saintillan \& Shelley \citep{saintillan2007orientational,saintillan2008instabilities} is recovered by setting $\phi=0$. 
%

\subsection{Results (nearly aligned)}

\begin{figure*}[p]
\includegraphics[width=\textwidth ,trim=0 1 0 0,clip]{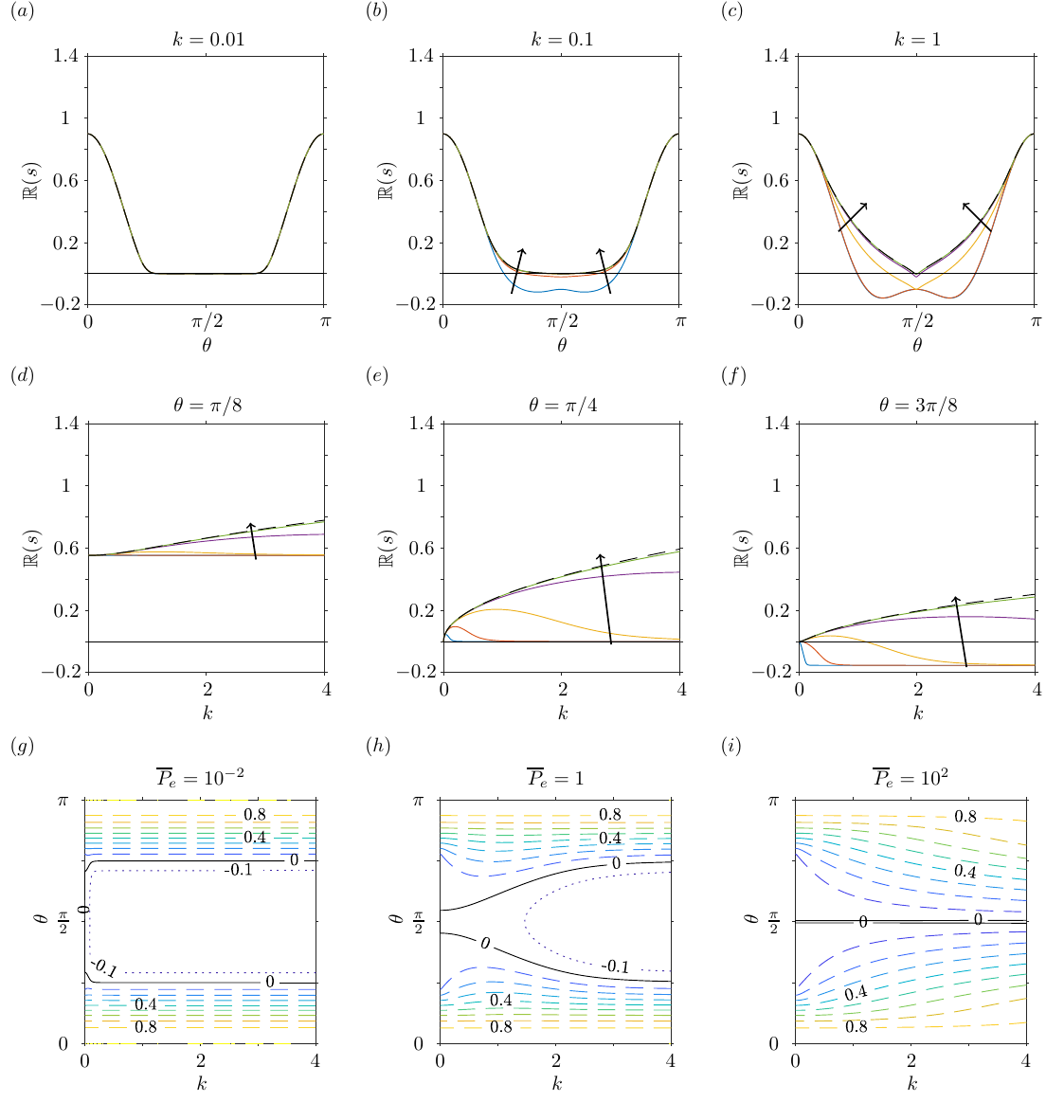}	
\caption{Linear stability analysis of a nearly aligned suspension, for variations in translational diffusivity (P\'{e}clet number) but neglecting anisotropic effects. In all plots the volume fraction, shape parameter and stresslet strength are constant ($\phi = 0.2$, $\alpha_0 = 0.8$, $\alpha_1=-1$).
Fig. $(a)$-$(f)$ show the real part of the growth rate ($\mathbb{R}(s)$) for changing P\'{e}clet number $\overline{P}\!_e=10^{-2},10^{-1},1,10^1,10^2$, where the arrow indicates the direction of increase. The dependence of $\mathbb{R}(s)$ on the wave-direction $\theta$ is shown for fixed values of the wave-number $(a)$ $k=0.01$, $(b)$ $k=0.1$, $(c)$ $k=1$, and the dependence of $\mathbb{R}(s)$ on $k$ for fixed values of $(d)$ $\theta = \pi/8$, $(e)$ $\theta = \pi/4$, $(f)$ $\theta =3 \pi/8$. The dashed line corresponds to $\phi=0$ and $1/\overline{P}\!_e=0$, \emph{i.e.}\ the diffusion-free regime considered by Saintillan \& Shelley \citep{saintillan2008instabilities}. Fig. $(g)$-$(i)$ show the dependence of $\mathbb{R}(s)$ on $k$ and $\theta$ for fixed values of $(g)$ $\overline{P}\!_e=10^{-2}$, $(h)$ $\overline{P}\!_e  =1$ and $(i)$ $\overline{P}\!_e=10^2$. Here the spacing of contour lines represents a change of $0.1$ to $\mathbb{R}(s)$. The black solid line in all plots indicates $\mathbb{R}(s)=0$, \emph{i.e.}\ the boundary between instability and stability.}\label{fig:NoAni}
\end{figure*}
This section will examine the growth rate of instability in a nearly aligned suspension of pushers ($\alpha<0$), given by equation \eqref{dispersion}, as translational diffusion, quantified by the P\'{e}clet number, volume fraction  and shape parameter are varied.

First, consider the case when there are no anisotropic effects (\emph{i.e.}\ $A_1=1$ and $A_2=1$), the volume fraction, shape parameter and stresslet strength are held constant ($\phi=0.2$, $\alpha_0=0.8$, $\alpha_1=-1$), and the P\'{e}clet number is varied ($\overline{P}\!_e = 10^{-2},10^{-1},1,10^1,10^2$). This variation in P\'{e}clet number corresponds to changes from large to small diffusion coefficient, with swimming velocity and length scale held constant. The corresponding results for this setup are shown in Fig. \ref{fig:NoAni}. 
When $k=0.01$, Fig. \ref{fig:NoAni}$(a)$ shows that changes to the P\'{e}clet number have no effect on the growth parameter $s$. However, for higher wave-numbers the inclusion of diffusion stabilises some wave-directions, for example in Fig. \ref{fig:NoAni}$(b)$-$(c)$,$(e)$-$(f)$. Diffusive effects may change the most unstable value of $k$. In Fig. \ref{fig:NoAni}$(d)$-$(f)$, when translational diffusion is sufficiently large, instead of growth rate being strictly increasing with $k$ there is now a finite positive maximum. Diffusion therefore has a stabilising effect, as predicted by ref. \citep{saintillan2007orientational,saintillan2008instabilities}. This effect is highlighted more clearly in Fig. \ref{fig:NoAni}$(g)$-$(i)$, in which the solid black lines separate the stable and unstable regions in $k$--$\theta$ space.

For small P\'{e}clet number there is a range of wave-directions ($\theta \in (\pi/4,3 \pi/4)$) for which $\mathbb{R}(s)<0$ for all wave-numbers; this can be identified by taking the small P\'{e}clet number limit of equation \eqref{dispersion}. However, there is always a range of wave-directions for which all wave-numbers are unstable ($\theta \in [0,\pi/4)$ and $\theta \in (3 \pi/4, \pi]$).
Therefore translational diffusion does not significantly dampen high wave-number instability (in contrast with the suggestion of ref.\ \citep{saintillan2008instabilities}). As P\'{e}clet number is increased (diffusion decreased) the range of stable wave-directions and corresponding wave-numbers is decreased (Fig. \ref{fig:NoAni}$(g)$-$(i)$), until the limit $1/\overline{P}\!_e =0$, corresponding to ref.\ \cite{saintillan2008instabilities}, is found where the perturbation is unconditionally unstable.

\begin{figure*}[p]
\includegraphics[width=\textwidth ,trim=0 1 0 0,clip]{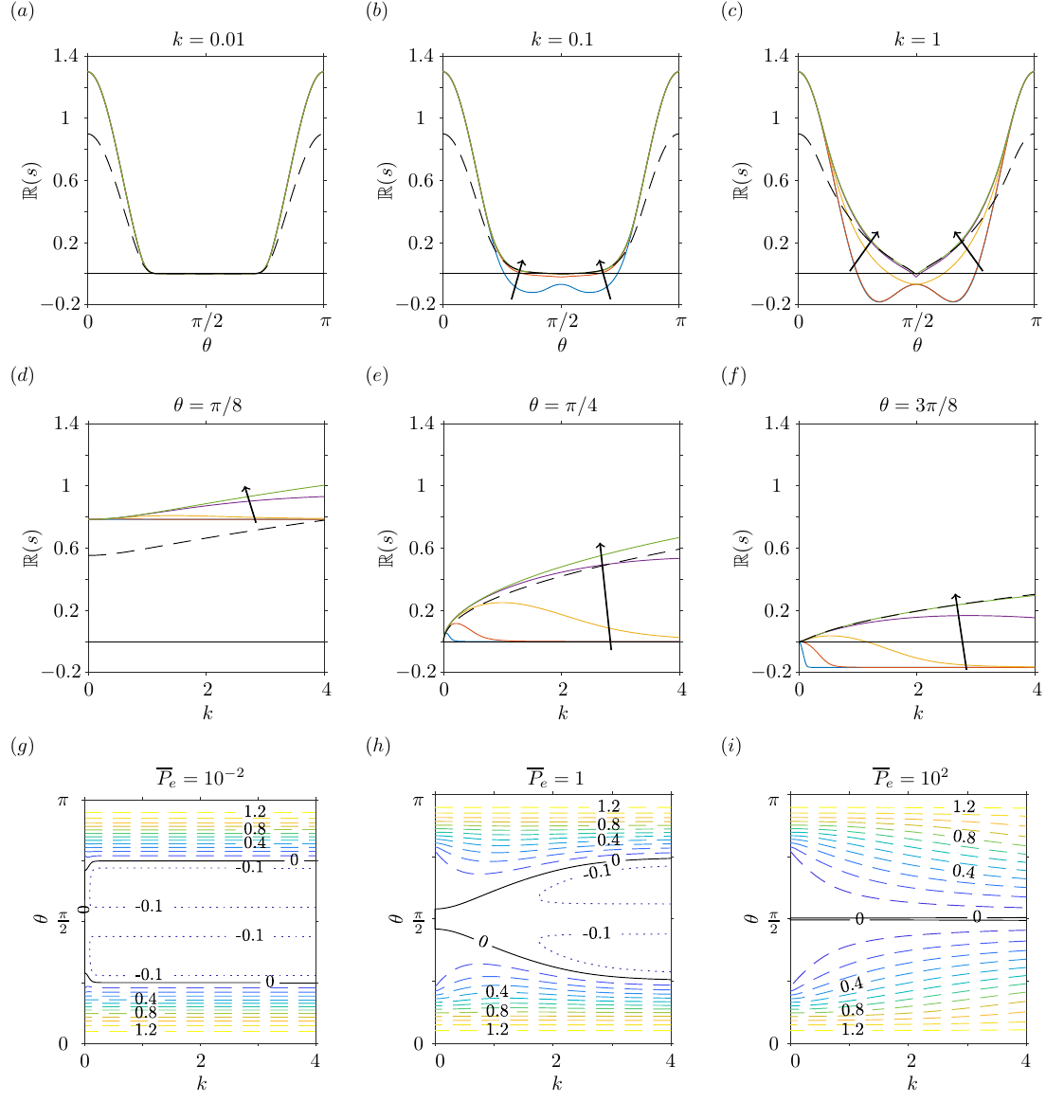}	
\caption{Linear stability analysis of a nearly aligned suspension, for variations in translational diffusivity (P\'{e}clet number) when mechanically anisotropic effects are included.
In all plots the volume fraction, shape parameter and stresslet strength are constant ($\phi = 0.2$, $\alpha_0 = 0.8$, $\alpha_1=-1$).
Fig. $(a)$-$(f)$ show the real part of the growth rate ($\mathbb{R}(s)$) for changing P\'{e}clet number $\overline{P}\!_e=10^{-2},10^{-1},1,10^1,10^2$, where the arrow indicates the direction of increase. The dependence of $\mathbb{R}(s)$ on the wave-direction $\theta$ is shown for fixed values of the wave-number $(a)$ $k=0.01$, $(b)$ $k=0.1$, $(c)$ $k=1$, and the dependence of $\mathbb{R}(s)$ on $k$ for fixed values of $(d)$ $\theta = \pi/8$, $(e)$ $\theta = \pi/4$, $(f)$ $\theta =3 \pi/8$. The dashed line corresponds to $\phi=0$ and $1/\overline{P}\!_e=0$, \emph{i.e.} the diffusion-free regime considered by Saintillan \& Shelley \citep{saintillan2008instabilities}. Fig. $(g)$-$(i)$ show the dependence of $\mathbb{R}(s)$ on $k$ and $\theta$ for fixed values of $(g)$ $\overline{P}\!_e=10^{-2}$, $(h)$ $\overline{P}\!_e  =1$ and $(i)$ $\overline{P}\!_e=10^2$. Here the spacing of contour lines represents a change of $0.1$ to $\mathbb{R}(s)$. The black solid line in all plots indicates $\mathbb{R}(s)=0$, \emph{i.e.}\ the boundary between instability and stability.}\label{fig:Ani}
\end{figure*}

In Fig. \ref{fig:Ani} we consider the effect of translational diffusion when anisotropic effects are included. As in Fig. \ref{fig:NoAni}, we choose $\phi=0.2$, $\alpha_0=0.8$ (\emph{i.e.}\ $\Gamma = 3$), $\alpha_1=-1$ and vary $\overline{P}\!_e = 10^{-2}, 10^{-1}, 1,$ $10, 10^2$, but now determine the values of $A_1$ and $A_2$ in equation \eqref{eq:47} from equations \eqref{eq:48}-\eqref{eq:49}, \eqref{eq:51}-\eqref{eq:58}.
We identify qualitatively similar results when anisotropic effects are included (Fig. \ref{fig:Ani}) as when they are neglected (Fig. \ref{fig:NoAni}). 
However, anisotropic effects increase the corresponding value of $\mathbb{R}(s)$ and therefore instabilities will grow more quickly (or decay more slowly). The boundary between stability and instability remains unchanged.

\begin{figure*}[h]
\includegraphics[width=\textwidth ,trim=0 1 0 0,clip]{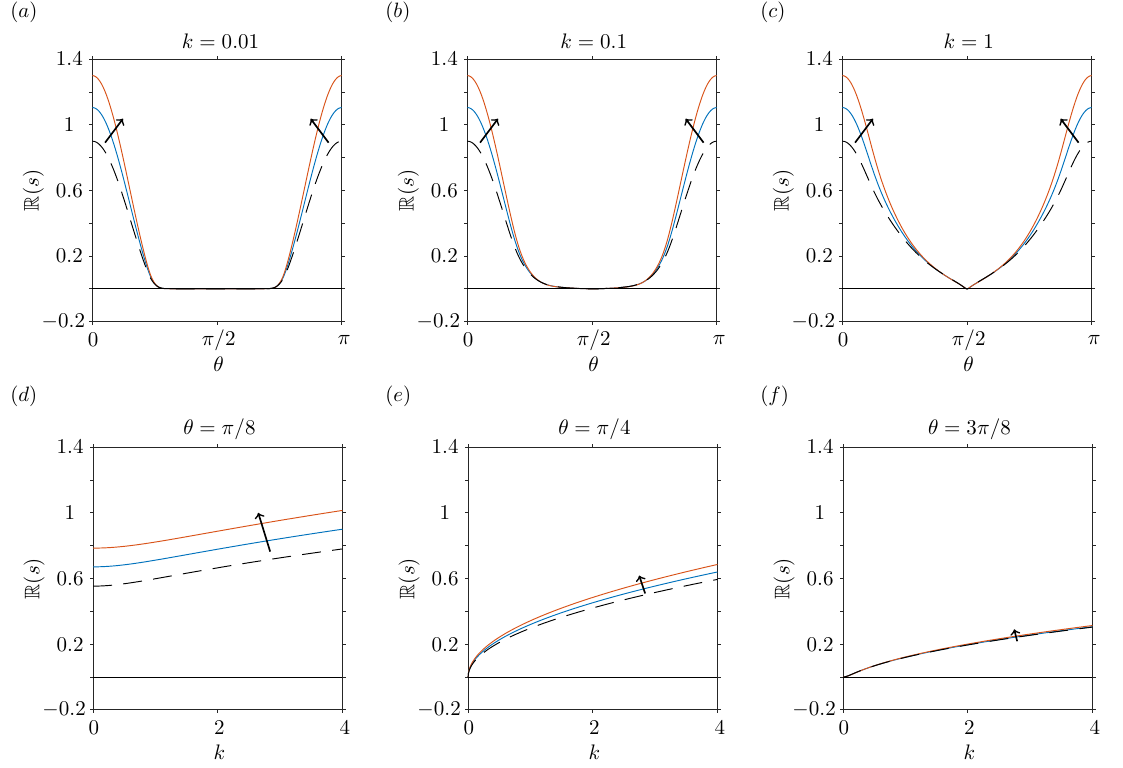}	
\caption{Linear stability analysis of a nearly aligned suspension, for variations in volume fraction ($\phi$), when mechanically anisotropic effects are included but translational diffusivity is neglected.
The real part of the growth rate ($\mathbb{R}(s)$) for changes in the volume fraction $\phi=0,0.1,0.2$, where the arrow indicates the direction of increase. The P\'{e}clet number, shape parameter and stresslet strength are constant ($1/\overline{P}\!_e=0$, $\alpha_0 = 0.8$, $\alpha_1=-1$). The dependence of $\mathbb{R}(s)$ on the wave-direction $\theta$ is shown for fixed values of the wave-number $(a)$ $k=0.01$, $(b)$ $k=0.1$, $(c)$ $k=1$, and the dependence of $\mathbb{R}(s)$ on $k$ for fixed values of $(d)$ $\theta = \pi/8$, $(e)$ $\theta = \pi/4$, $(f)$ $\theta =3 \pi/8$. The dashed line represents the case $\phi=0$, \emph{i.e.}\ the regime considered by Saintillan \& Shelley \citep{saintillan2008instabilities}. The black solid line in all plots indicates $\mathbb{R}(s)=0$, \emph{i.e.}\ the boundary between instability and stability.}\label{fig:phi}
\end{figure*}
\begin{figure*}[h]
\includegraphics[width=\textwidth ,trim=0 1 0 0,clip]{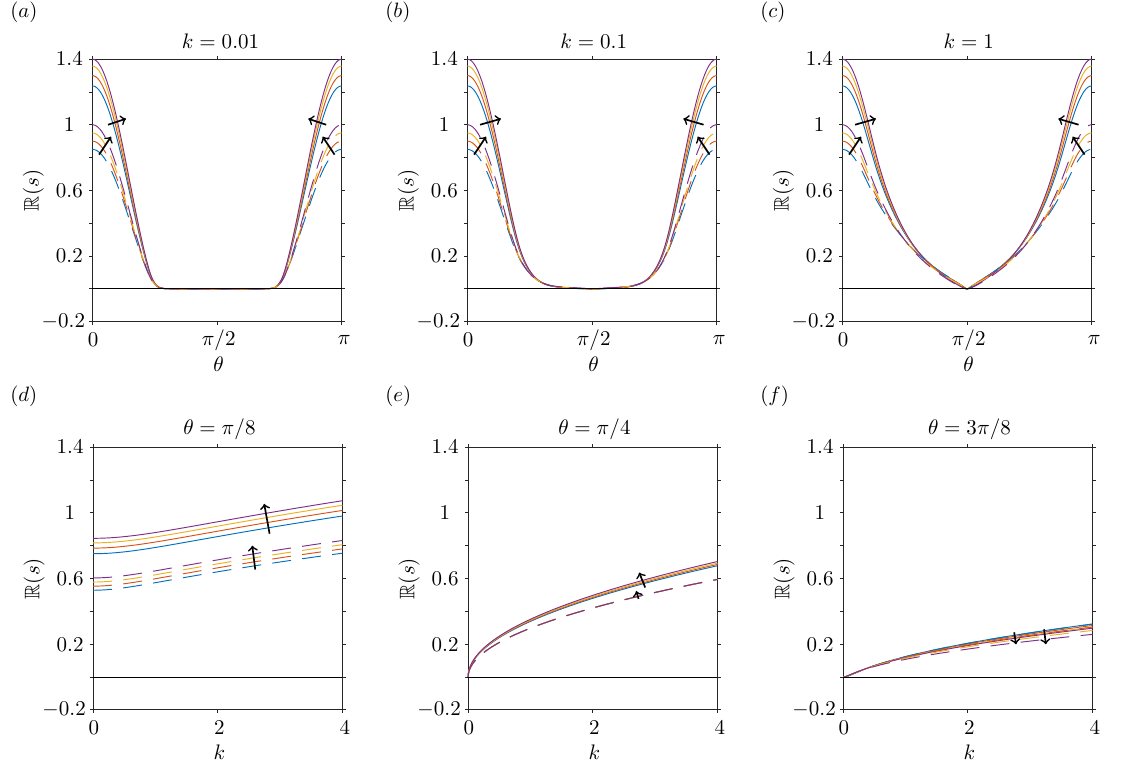}	
\caption{Linear stability analysis of a nearly aligned suspension, for variations in shape parameter ($\alpha_0$), when mechanically anisotropic effects are included but translational diffusivity is neglected.
The real part of the growth rate ($\mathbb{R}(s)$) for changing shape parameter $\alpha_0=0.7,0.8,0.9,1$, where the arrow indicates the direction of increasing $\alpha_0$. The solid (dashed) lines indicate when transversely-isotropic effects are (not) included. The P\'{e}clet number, volume fraction and stresslet strength are constant ($\overline{P}\!_e \gg 1$, $\phi=0.2$, $\alpha_1=-1$). The dependence of $\mathbb{R}(s)$ on the wave-direction $\theta$ is shown for fixed values of the wave-number $(a)$ $k=0.01$, $(b)$ $k=0.1$, $(c)$ $k=1$, and the dependence of $\mathbb{R}(s)$ on $k$ for fixed values of $(d)$ $\theta = \pi/8$, $(e)$ $\theta = \pi/4$, $(f)$ $\theta =3 \pi/8$. The black solid line in all plots indicates $\mathbb{R}(s)=0$, \emph{i.e.}\ the boundary between instability and stability.}\label{fig:alpha0}
\end{figure*}

In Fig. \ref{fig:phi} we identify the importance of the volume fraction $\phi$ when diffusion is negligible ($1/\overline{P}\!_e =0$), $\alpha_0=0.8$, $\alpha_1=-1$ and anisotropic effects are included. The growth curves are shown for $\phi =0,0.1,0.2$.
We observe in Fig. \ref{fig:phi}$(a)$-$(c)$ there is always a positive growth rate; therefore, aligned suspensions are always unstable to concentration and orientation perturbations. This agrees with the results presented by Saintillan \& Shelley \citep{saintillan2007orientational,saintillan2008instabilities} which correspond to $\phi=0$ (the dashed line in Fig. \ref{fig:phi}). However, our model predicts that the perturbations will grow more quickly as the volume fraction of particles is increased. From Fig. \ref{fig:phi}$(d)$-$(f)$ we note the largest change to the growth rate, with respect to change in volume fraction, occurs when the wave-angle $\theta = \pi/8$, and the least for $\theta =3\pi/8$. Therefore the inclusion of the extra stress $\bm\sigma\!_P$ has the greatest effect on instabilities with wave-direction similar to the fibre alignment, and little to no effect on instabilities with wave-direction perpendicular to the fibre direction. 
For the parameter range considered, anisotropic effects more pronounced than shape effects (Fig. \ref{fig:alpha0}).

\subsection{Stability of an isotropic suspension}
We now examine the stability of a suspension of randomly orientated particles by perturbing a uniform steady state, where the background fluid is stationary:
\begin{align}
{\bm u} (\bm{x},t)&= \varepsilon {\bm u}^{(1)}({\bm x},t) + \mathcal{O} \left( \varepsilon^2 \right), \nonumber\\ N(\bm{x},\hatp,t) &= \frac{1}{4\pi} \left[ 1 + \varepsilon N^{(1)}({\bm x}, \hatp,t)\right] + \mathcal{O} \left( \varepsilon^2 \right), \nonumber \\
p(\bm{x},t) &= p_0 + \varepsilon \, p^{(1)}(\bm{x},t) + \mathcal{O} \left( \varepsilon^2 \right). \label{AIeq:iso}
\end{align} 
The following quantities may then be found
\begin{align}
\begin{split}
\langle \hatp & \rangle = \bm{0}, \qquad	 \langle \hatp \, \hatp \rangle = \frac{\bm{ I}}{3}, \\{\bm e}^{(1)} &: \int_S \hatp \, \hatp  \, \hatp \, \hatp \dif \hatp = \frac{2}{15} {\bm e}^{(1)}, \label{relations}
\end{split}
\end{align}
where $\bm{e}^{(1)}=(\gradx {\bm u}^{(1)} +\gradx {\bm u}^{(1)T})/2$ is the rate-of-strain tensor at order $\varepsilon$ and we have used that the integrals are isotropic \citep{dyson2015investigation,spain1953tensor}.

Using the relations \eqref{relations} the Fokker-Planck equation \eqref{eq:N} may be simplified to 
\begin{align}
\dpd{N^{(1)}}{t} &= - \hatp \cdot \gradx N^{(1)} + \frac{\phi}{\overline{P}\!_e} \nabla_{\bm x}^2 N^{(1)} + \underbrace{5 \alpha_0 \hatp \, \hatp : {\bm e}^{(1)}}_{(*)}. \label{AIeq:isoN}
\end{align}
We note the final term $(*)$ differs from Saintillan \& Shelley \cite{saintillan2008instabilities}, due to an error made when calculating $\delta_{ii}$ (Appendix \ref{app:SaintillanError}).

Using equation \eqref{relations} and the constitutive equation for stress \eqref{stress} the governing equation for the flow velocity \eqref{eq:flow} becomes
\begin{align}
\gradx \cdot {\bm u}^{(1)} &= 0, \\ - \tilde{\mu} \nabla_{\! \bm{x}}^2 {\bm u}^{(1)} + \gradx p^{(1)} &= \frac{\alpha_1}{4 \pi} \int_S \left( \hatp \, \hatp - \frac{\bm I}{3} \right) N^{(1)} \dif \hatp,  \label{AIeq:flow_1}
\end{align}
where $\tilde{\mu} = 1 + 4 \phi (2 \alpha_2 / 15 + 2 \alpha_3 /3 + \alpha_4)$.
We note that equation \eqref{AIeq:flow_1} is identical to that presented by Saintillan \& Shelley \citep{saintillan2007orientational,saintillan2008instabilities} by setting $\phi=0$.

A dispersion relation for the growth parameter $s$ may be derived by following a similar method to ref.\ \citep{saintillan2008instabilities}, we summarise the steps here.
We look for plane-wave perturbations of the form 
\begin{align}
N^{(1)} &= \tilde{N}({\bm k},\hatp) e^{i {\bm k} \cdot {\bm x} + s t}, \label{64} \\ {\bm u}^{(1)} &= \tilde{\bm u}({\bm k}) e^{i {\bm k} \cdot {\bm x} + s t}, \\ p^{(1)} &= \tilde{p}({\bm k}) e^{i {\bm k} \cdot {\bm x} + s t},\\ {\bm e}^{(1)} &= \tilde{\bm e}({\bm k}) e^{i {\bm k} \cdot {\bm x} + s t}, \\
{\bm{\sigma}}^{(1)}_S &= \tilde{\bm{\sigma}}_S({\bm k}) e^{i {\bm k} \cdot {\bm x} + s t}, \label{AIeq:iso_boundary}
\end{align}
where ${\bm k}$ is the wave-vector and $s$ is the growth rate.
The rate of strain tensor may then be evaluated as
\begin{align}
\tilde{\bm e} &= \frac{i}{2} \left( \tilde{\bm u} \, {\bm k} + {\bm k} \, \tilde{\bm u} \right),\label{TildeStrain}
\end{align}
where
\begin{align}
\tilde{\bm u} &= \frac{i}{k \tilde{\mu}} \left( \bm{I} - \hatk \, \hatk \right) \cdot \tilde{\bm \sigma}_S \cdot \hatk, \label{68}\\
\tilde{\bm \sigma}_S &= \frac{\alpha_1}{4 \pi} \int_S \left( \hatp \, \hatp - \frac{\bm I}{3} \right) \tilde{N} \dif \hatp,	\label{69}
\end{align}
and $\hatk = \bm{k}/k$ where $k=|\bm{k}|$.
Substituting equations \eqref{68} and \eqref{69} into equation \eqref{TildeStrain}, using the fact that $\hatk$ is a unit vector and finally substituting the ansatzes into equation \eqref{AIeq:isoN} yields the eigenvalue relation
\begin{align}
\bm{F} [ \tilde{N} ] &= \frac{- 5 \alpha_0 \alpha_1 }{4 \pi \tilde{\mu}} \int_S \frac{( \hatk \cdot \hatp' )^2 ( \bm{I} - \hatk \, \hatk ) \cdot \hatp' \, \hatp' \cdot {\bf F} [\tilde{N}]}{s + i \bm{k} \cdot \hatp' + \phi k^2/\overline{P}\!_e} \dif \hatp',	  \label{eigRelation}
\end{align}
where the operator $\bm{F}$ is defined as
\begin{align}
\bm{F}[\tilde{N}] &= \left( \bm{I} - \hatk \, \hatk \right) \cdot \int_S \hatp' \left( \hatp' \cdot \hatk \right) \tilde{N} \dif \hatp'.
\end{align}
Note $\bm{F}[\tilde{N}]$ has its $\hatp$ dependence integrated out and therefore cancels on both side of equation \eqref{eigRelation}.
By noting the eigenvalue relation \eqref{eigRelation} is invariant under rotation we may choose $\hatk=\hatz$ without loss of generality. After evaluating the surface integrals the dispersion relation is given as 
\begin{align}
\frac{ 5 i \alpha_0 \alpha_1}{4 k \tilde{\mu}} \left[ 2 \tilde{\lambda}^3 - \frac{4}{3} \tilde{\lambda} + \left( \tilde{\lambda}^4 - \tilde{\lambda}^2 \right) \log \left( \frac{ \tilde{\lambda}-1}{\tilde{\lambda}+1} \right) \right] &= 1,\label{AIeq:dispersion}
\end{align}
where $\tilde{\lambda}=-i(s + \phi k^2/\overline{P}\!_e)$. 
Equation \eqref{AIeq:dispersion} differs from that found by Saintillan \& Shelley \citep{saintillan2008instabilities} through the additional anisotropic contribution to the viscosity $\tilde{\mu}$, and the leading numerical factor, which arises from correction discussed in appendix \ref{app:SaintillanError}.
Equation \eqref{AIeq:dispersion} is a dispersion relation for the growth rate $s$ (via $\tilde{\lambda}$) and may be solved numerically using Newton's method.

\subsection{Results (nearly isotropic)}
This section will examine the growth rate of instability in a nearly isotropic suspension of pushers ($\alpha_1 <0$), given by equation \eqref{AIeq:dispersion}, as the volume fraction and shape parameter are varied.
\begin{figure*}[h]
\includegraphics[width=\textwidth ,trim=0 1 0 0,clip]{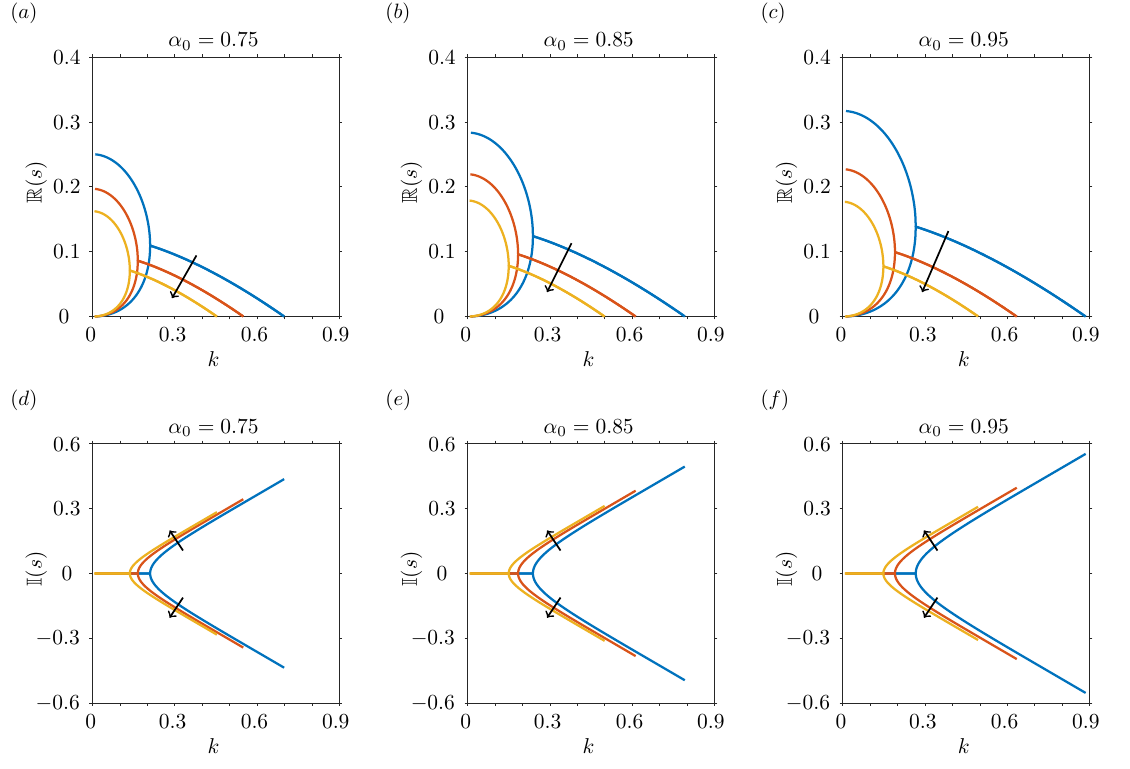}	
\caption{Linear stability analysis of a nearly isotropic suspension, for variations in volume fraction ($\phi$) and shape parameter ($\alpha_0$), when mechanically anisotropic effects are included but translational diffusivity is neglected. Note that in the limit $\phi \rightarrow 0$, this does not agree quantitatively with Saintillan \& Shelley \citep{saintillan2008instabilities} due to the numerical prefactor correction in equation \eqref{AIeq:dispersion}. The real ($a$-$c$) and imaginary ($d$-$f$) parts of the dispersion relation for changing volume fraction $\phi = 0, 0.05, 0.1$, where the arrows shows the direction of increase, for fixed shape parameters $(a),(d)$ $\alpha_0 =0.75$, $(b),(e)$ $\alpha_0 =0.85$, $(c),(f)$ $\alpha_0 =0.95$.}\label{fig:iso}
\end{figure*}

Fig. \ref{fig:iso} shows the real and imaginary parts of the growth parameter $s$, as a function of wave-number $k$ for selected values of the shape parameter $\alpha_0$ and volume fraction $\phi$. 
We observe for small wave-numbers the growth rate is real and positive, therefore small wave-number disturbances will grow exponentially in suspensions of pushers.
For higher wave-numbers the two branches of the growth parameter become a complex conjugate pair with $\mathbb{R}(s)>0$, implying that any disturbances will grow and also oscillate. 

As the volume fraction is increased the growth rate decreases for corresponding wave-numbers, but will oscillate more as the imaginary part of $s$ increases. We observe that the plot of the dispersion relation takes the same form, becoming dampened as $\phi$ increases. This means a smaller range of wave-numbers will become unstable for each particle size.

\section{Conclusion}\label{summary}
In this paper we have linked models of active suspensions of elongated motile particles to the transversely-isotropic fluid proposed by Ericksen \citep{ericksen1960transversely}, using a Fokker-Planck equation for the particle distribution function and the stress tensor of Pedley \& Kessler \citep{pedley1990new}, which includes the influence of non-spherical particles on the bulk stress. Under the assumption that the suspension is of spatially uniform volume fraction and has perfect but perhaps spatially varying alignment, Ericksen's four viscosity-like parameters may be determined in terms of fundamental physical quantities. These quantities include the active stresslet, particle aspect ratio, particle volume, mean number density of the particles and solvent viscosity. This linkage yields a physical basis for inferring these crucial mechanical parameters used in models such as \citep{green2008extensional,dyson2015investigation,lee2005continuum}. The shear-independent term parameterised by $\mu_1$ is found to model active behaviour. The transversely-isotropic fluid of Ericksen may therefore be used to model actively motile `fibres' by a simple modification to the fibre evolution equation. Linking these two frameworks provides a basis to extend Ericksen's model to include effects such as dispersion about the preferred direction.

Our modification to Ericksen's model can be considered as the simplest describing an orientated active suspension and including transversely-isotropic effects; more refined approaches take into account fibre-dispersion formulated via the $Q$-tensor which is defined as the nematic moment $\langle \bm{p\,p} - \bm{I}/3\rangle$ \citep{woodhouse2012spontaneous,brotto2013hydrodynamics}.

Motivated by this linkage between transversely-isotropic fluids and active suspension modelling, we examined the linear stability of the active suspension in two special cases, the first when the particles are uniformly distributed and perfectly aligned, and the second when the particles are uniformly randomly orientated (an isotropic suspension), both with no imposed background flow. We found the results of Saintillan \& Shelley \citep{saintillan2007orientational,saintillan2008instabilities} are the zero-volume-fraction limit of the model, up to a numerical factor.

To investigate the linear stability of an aligned suspension we found a base state, in which the distribution is similar to that of a transversely-isotropic fluid, except with non-constant distribution of particles. We then assumed the perturbation to the base state was of the form of a plane wave, and noted that the first order velocity was only non-zero when the wave-vector lay in the $(\hatz, \bm{a}')$-plane. Here $\hatz$ is the direction of the base state alignment whilst $\bm{a}'$ is the alignment of the first order perturbation. We hence found the dispersion relation, and identified the range of wave-vectors which are unstable for suspensions of pusher particles. 
If translational diffusion is neglected, the growth rate increases with wave-number, as predicted by Saintillan \& Shelley \citep{saintillan2007orientational,saintillan2008instabilities}.  
Even in the presence of translational diffusion, some wave-directions remain unstable for all wave-numbers, \emph{i.e.}\ diffusion selects a bounded range of unstable wave-numbers.
Anisotropic effects do not change the unstable wave-vectors, however they do increase the growth rate of perturbations. 

Considering an initially randomly oriented (isotropic) suspension at rest, the base state corresponds to a constant particle distribution function and zero fluid velocity, giving isotropic integrals for the first, second and fourth moments of the distribution function, which may be evaluated analytically.
By following a similar method to Saintillan \& Shelley \citep{saintillan2007orientational,saintillan2008instabilities} we found a dispersion relation for the growth rate.
 For a suspension of pushers, low wave-number perturbations ($k \in [0,0.15]$) grow exponentially with no oscillations, whilst medium wave-number ($k \in [0.15,0.6]$) perturbations oscillate and grow more slowly. Once the wave-number is large enough perturbations are dampened. The inclusion of the extra stress due to interactions between elongated particles and the surrounding fluid simply decreases the growth rate for corresponding wave-numbers as the volume fraction of particles is increased. This leads to a smaller range of wave-numbers when the perturbation is unstable when compared to results for an isotropic stress tensor.

Active suspensions and transversely-isotropic fluids are both biologically-relevant and physically-fascinating states of matter. Linking these two fields of research enables knowledge transfer, enabling extension of the transversely-isotropic model and identifying important components of the active suspension model.
Transversely-isotropic effects were found to produce some changes to linear stability analysis of nearly-aligned and isotropic active suspensions in the absence of background flow. We argue that these aspects should be included in future modelling studies.

\begin{acknowledgement}
{\bf Acknowledgements} \\
CRH is supported by an Engineering and Physical Sciences Research Council (EPSRC) doctoral training award (EP/J500367/1), GC is supported by a Biotechnology and Biological Sciences Research Council Industrial CASE studentship (BB/L015587/1), JEFG acknowledges the support of the Australian Research Council (ARC) Discovery Early Career Researcher Award (DE130100031) and RJD the support of the EPSRC grant (EP/M00015X/1). CRH is grateful to  Universitas 21 and the Institute of Mathematics and its Applications for supporting travel to visit JEFG and RJC.
The authors thank Dr Louise Dyson, University of Warwick, for helpful discussions.
CRH, GC and RJD would like to acknowledge the hospitality of the University of Auckland and University of Adelaide and JEFG and RJC that of the University of Birmingham during research and study leave.
\end{acknowledgement}

\noindent
{\bf Author Contributions}\\
CRH led the development of the mathematical model, with intellectual contributions from all authors. RJD supervised the project and designed the research with contributions from all authors. CRH, DJS and RJD wrote the paper with contributions from all authors.


\begin{appendices}
\section{Viscosity parameters for a dilute suspension}\label{app:Jeffery}
For a dilute suspension the parameters $\mu$, $\alpha_2$ and $\alpha_3$ may be approximated from Jeffery \citep{jeffery1922motion} (via \citep{pedley1990new,batchelor1970stress,leal1971effect,brenner1972rheology,kim2013microhydrodynamics} 
\begin{align}
\mu &=  1 + \frac{2 \phi}{I_1} , \label{eq:51} \\ \alpha_2 &= \frac{1}{I_1} \left( 1 + \frac{L_1}{L_2} - 2 \frac{I_1}{I_2} \right), \\ \alpha_3 &= \frac{1}{I_1} \left( \frac{I_1}{I_2} - 1 \right).
\end{align}
The quantities $I_1$, $I_2$, $L_1$ and $L_2$ are ellipsoidal integrals, given in terms of the aspect ratio $\Gamma$:
\begin{align}
I_1 &= \frac{\Gamma^2 \left( 2 \Gamma^2 - 5 + 3 \gamma \right)}{2 (\Gamma^2 -1)^2}, \\ I_2 &= \frac{\left( \Gamma^2 + 1 \right) \left( \Gamma^2 + 2 - 3 \Gamma^2 \gamma \right)}{(\Gamma^2-1)^2}, \\
L_1 &= \frac{ \Gamma^2 \left[ 2 \Gamma^2 + 1 - \gamma \left( 4 \Gamma^2 - 1 \right)\right]}{4 (\Gamma^2 -1)^2}, \\ L_2 &= I_1 - 2 L_1,  \\
\gamma &= \frac{ \cosh^{-1} \Gamma}{\Gamma (\Gamma^2 - 1)^{1/2}}. \label{eq:58}
\end{align}

\section{Derivation of first order Fokker-Planck equation}\label{app:SaintillanError}	
Equation \eqref{eq:N} may be simplified by noting the rate-of-strain and vorticity tensors are zero at leading order and the fluid is incompressible,
\begin{align}
\begin{split}
\dpd{N^{(1)}}{t} =& - \gradx \cdot \left( \hatp N^{(1)} \right) + \frac{\phi}{\overline{P}\!_e} \nabla_{\! \bm{x}}^2 N^{(1)} \\ 
&- \gradp \cdot \left[  \left( \bm{I} - \hatp \, \hatp \right) \cdot \left( \alpha_0 \bm{e}^{(1)} + \bm{\omega}^{(1)} \right) \cdot \hatp \right]. \label{eq:568}
\end{split}
\end{align}
The final term of equation \eqref{eq:568} may be simplified further, to see this we use index notation
\begin{align}
\gradp \cdot & \Bigg[  \left( \bm{I} - \hatp \, \hatp \right)  \cdot \left( \alpha_0 \bm{e}^{(1)} + \bm{\omega}^{(1)} \right) \cdot \hatp \Bigg] \nonumber \\
&= \dpd{}{\hat{p}_i} \left[ \left( \delta_{ij} - \hat{p}_i \, \hat{p}_j \right) \left( \alpha_0 e_{jk}^{(1)} + \omega_{jk}^{(1)} \right) p_k \right],  \nonumber   \\
&= - \left( \alpha_0 e_{ik}^{(1)} + \omega_{ik}^{(1)} \right) \dpd{\hat{p}_k}{\hat{p}_i} + \alpha_0 e_{jk}^{(1)} \dpd{}{\hat{p}_i} \left( \hat{p}_i \hat{p}_j \hat{p}_k \right) \nonumber \\
& \hspace{1cm} + \dpd{}{p_i} \left( \hat{p}_i \left[  \hat{p}_j \hat{p}_k \omega_{jk}^{(1)} \right] \right).   \label{eq:569}
\end{align}
The first term in equation \eqref{eq:569} is zero via the incompressibility condition \eqref{eq:flow} and since the vorticity tensor is traceless, \emph{i.e.}\
\begin{align}
\left( \alpha_0 e_{ik}^{(1)} + \omega_{ik}^{(1)} \right) \dpd{\hat{p}_k}{\hat{p}_i} &= \left( \alpha_0 e_{ik}^{(1)} + \omega_{ik}^{(1)} \right) \delta_{ki}, \nonumber \\
&= \alpha_0 e_{ii}^{(1)} + \omega_{ii}^{(1)}, \nonumber \\
&= \alpha_0 \dpd{u_i^{(1)}}{x_i} + \frac{1}{2} \left( \dpd{u_i^{(1)}}{x_i} - \dpd{u_i^{(1)}}{x_i} \right), \nonumber \\
&= 0.
\end{align}
The final term is also zero, to see this we write the vorticity tensor in terms of velocity gradients and simplify:
\begin{align}
2 \hat{p}_j \hat{p}_k \omega_{jk} &= \dpd{u_j^{(1)}}{x_k} \hat{p}_j \hat{p}_k - \dpd{u_k^{(1)}}{x_j} \hat{p}_j \hat{p}_k, \nonumber \\
&= \dpd{u_j^{(1)}}{x_k} \hat{p}_j \hat{p}_k - \dpd{u_j^{(1)}}{x_k} \hat{p}_k \hat{p}_j, \nonumber \\
&=0.
\end{align}
Therefore equation \eqref{eq:569} is given by 
\begin{align}
\gradp \cdot & \Bigg[  \left( \bm{I} - \hatp \, \hatp \right)  \cdot \left( \alpha_0 \bm{e}^{(1)} + \bm{\omega}^{(1)} \right) \cdot \hatp \Bigg] \nonumber \\
&= \alpha_0 e_{jk}^{(1)} \dpd{}{\hat{p}_i} \left( \hat{p}_i \hat{p}_j \hat{p}_k \right), \nonumber \\
&= \alpha_0 e_{jk}^{(1)} \left[ \hat{p}_i \hat{p}_j \delta_{ki} + \hat{p}_i \hat{p}_k \delta_{ji} + \hat{p}_j \hat{p}_k \delta_{ii} \right], \nonumber \\
&= \alpha_0 e_{jk}^{(1)} \left[ \hat{p}_j \hat{p}_k + \hat{p}_j \hat{p}_k + 3 \hat{p}_j \hat{p}_k \right] \nonumber \\
&= 5 \alpha_0 \bm{e}^{(1)} : \hatp \, \hatp. \label{eq:578}
\end{align}
Substituting equation \eqref{eq:578} into equation \eqref{eq:569} gives the first order Fokker-Planck equation
\begin{align}
\dpd{N^{(1)}}{t} &= - \hatp \cdot \gradx N^{(1)} + \frac{\phi}{\overline{P}\!_e} \nabla_{\bm x}^2 N^{(1)} + \underbrace{5 \alpha_0 \hatp \, \hatp : {\bm e}^{(1)}}_{(*)}. \label{AIeq:isoN}
\end{align}
We note the final term $(*)$ differs from Saintillan \& Shelley \cite{saintillan2008instabilities}, since $\delta_{ii}=3$ when summed over all indices.

\end{appendices}

\end{document}